\documentstyle[12pt,a4wide]{article}
\def\lesssim{\mathrel{\mathpalette\vereq<}}
\def\gtrsim{\mathrel{\mathpalette\vereq>}}
\def\vereq#1#2{\lower3pt\vbox{\baselineskip1.5pt \lineskip1.5pt
               \ialign{$\mpth#1\hfill##\hfil$\crcr#2\crcr\sim\crcr}}}
\def\mpth{\mathsurround=0pt}
\begin{document}

\begin{center}
\def\thefootnote{\fnsymbol{footnote}}
\Large
NEUTRINO MASSES AND MIXING\footnote{
Presented at the II$^{\mathrm{nd}}$ Rencontres du Vietnam Conference
"At the Frontiers of the Standard Model",
Ho Chi Minh City, Vietnam, October 1995.
}
\\
\vspace{0.5cm}
\large
S.M. Bilenky
\\
\it Joint Institute for Nuclear Research, Dubna, Russia\\
\it and\\
\it Scuola Internazionale Superiore di Studi Avanzati,\\
\it I-34013 Trieste, Italy
\end{center}

\vspace{0.5cm}
 
The investigation of neutrino properties is a very important problem
of today's  neutrino physics. The key problem is the problem of
neutrino  masses. If neutrinos are massive, they can be mixed. If the
total lepton number is conserved, massive neutrinos are Dirac
particles. If the neutrino masses  are generated by an  interaction
that does not conserve the total lepton number, neutrinos with
definite mass are Majorana particles. All this possibilities
correspond to different gauge  theories and it is very important that
they can be distinguished experimentally. There are more than 60
different experiments that are going on at present on the search for
effects  of the masses, nature and mixing of neutrinos, and there is a
general belief that investigation of neutrino properties could allow
us to reach new  physics.

In this reports I will discuss:

\noindent
1. possibilities of neutrino mixing.

\noindent
2. neutrino oscillations in vacuum and in matter.

\noindent
3. experimental indications in favor of neutrino masses and mixing.

From  all existing data it follows that flavor neutrinos 
$\nu_e,\nu_\mu$ and $\nu_\tau$
interact with matter via standard $CC$ and $NC$ interactions
\begin{equation}
{\cal{L}^{CC}}=-\frac{g}{2\sqrt 2}{j}^{CC}_{\alpha}W^{\alpha}+h.c.
\quad , \qquad \qquad
{\cal{L}^{NC}}=
-\frac{g}{2\cos \theta_{W}}{j}_{\alpha}^{NC}Z^{\alpha}
\quad ,
\label{eq:1}
\end{equation}
where 
\begin{equation}
j_{\alpha}^{CC}=2\sum_{l=e,\mu,\tau }\bar{\nu}_{lL}\gamma_{\alpha}l_L+...
\quad, \qquad \qquad
j_{\alpha}^{NC}=\sum_{l=e,\mu,\tau }\bar{\nu}_{lL}\gamma_{\alpha}\nu_{lL}+... 
\label{eq:2}
\end{equation}
Let us notice that the CC interaction
determines the concept of flavor 
neutrinos. For example, we call $\nu_\mu$ the particle that
 is produced together
with $\mu^{+}$ in $\pi^{+}\to {\mu^{+}\nu_{\mu}}$ decay, and so on.
According to the neutrino mixing hypothesis,
a flavor neutrino field
$\nu_{lL}$ is a unitary superposition of left-handed components of 
fields of neutrinos with definite masses:
\begin{eqnarray}
\nu_{lL}=\sum_{i} U_{li}\nu_{iL} ,\hskip2cm l=e,\mu,\tau
\end{eqnarray}
where $\nu_{i}$ is the field of the neutrino
with mass $m_i$ and $U$ is a unitary 
matrix.

The type of neutrino mixing is determined by the type of neutrino
mass term. There are three possible neutrino mass terms \cite{Bilenky}.
In the case of the Dirac 
mass term
\begin{eqnarray}
{\cal{L}}^{D} = -\sum_{l',l}{\bar\nu_{l',R}} M_{l'l} \nu_{lL} + h.c.
\end{eqnarray}
three neutrinos with definite masses are Dirac particles.
The Dirac mass 
term can be generated by the standard Higgs
mechanism. If this mass term enters in
the Lagrangian, the total lepton number
\begin{equation}
L=L_e+L_{\mu}+L_{\tau}
\end{equation}
is conserved.

There are two possible neutrino mass terms that do not conserve L:                       

\noindent
1.
The left-handed Majorana mass term
\begin{equation}
{\cal{L}}_{L}^{M}=-\frac{1}{2}\sum_{l',l}\overline{(\nu_{l'L})}^{c}
M_{l'l}^{L}\nu_{lL} + h.c.
\label{07}
\end{equation}
where $(\nu_{lL})^c=C\bar\nu_{lL}^T $
is the charge conjugated spinor. 
This mass term can be generated in models with a Higgs triplet. Let us notice
that theory with ${\cal{L}}^{M}_{L} $ is the most economical theory of
massive neutrinos (only left-handed fields enter in the Lagrangian).
From Eq.(\ref{07}),
for the flavor neutrino field we have
\begin{equation}
\nu_{lL}=\sum^{3}_{i=1} U_{li}\chi_{iL}
\end{equation}
where $ \chi_{i}=\chi_{i}^{c}\equiv C\bar\chi_i^T $ is the field of a 
Majorana neutrino with mass $m_i $.

\noindent
2.
The Dirac and Majorana mass term
\begin{equation}
{\cal{L}}^{D+M}={\cal{L}}^{M}_{L}+{\cal{L}}^D+{\cal{L}}^{M}_R
\end{equation}
This is the most general neutrino mass term that does not conserve $L$
and include both $\nu_{lL} $ and $\nu_{lR}$.
Dirac and Majorana mass term is typical for GUT models. For the mixing we 
have  
\begin{equation}
\nu_{lL}=\sum_{i=1}^{6} U_{li}\chi_{iL}
\quad , \qquad \qquad
(\nu_{lR})^c=\sum_{i=1}^{6} U_{\bar{l}i}\chi_{iL}
\quad ,
\label{10}
\end{equation}
where $\chi_{i}$ $(i=1,...,6)$
is the field of a Majorana particle with mass $m_i$ and 
$ U $ is a $ 6 \times 6 $ mixing matrix.
 
In the framework of the models with a Dirac and Majorana mass term,
exists the most popular mechanism of neutrino mass generation, the so
called see-saw mechanism \cite{seesaw}.
Assume that $ {\cal{L}}_{L}^{M}=0 $, $ {\cal{L}}^D $
is characterized by parameters that are
of the order of the fermion masses and
$ {\cal{L}}_{R}^{M} $ is characterized by parameters that are  of the 
order of $M_{GUT} $.
In this case, in the spectrum of the six Majorana 
particles there are three neutrinos with small masses
\begin{equation}
m_i\simeq \frac{(m_{F}^{i})^2}{M_i} \hskip2cm (i=1,2.3)
\end{equation}
and three particles with very heavy masses $M_i \simeq {M_{GUT}}$.
Here $ m_{F}^i $ is the mass of the $up$-quark or charged lepton in the
corresponding generation.
The see-saw mechanism provides 
a natural explanation of the smallness of neutrino masses.

If all the masses of Majorana particles in (\ref{10}) are small, transitions
of flavor neutrino into sterile states
$ \nu_l \to \bar{\nu}_{l'L} $
become possible
($\bar{\nu}_{lL}$ is a left-handed antineutrino,
a quantum of the right-handed field $\nu_{lR}$).

After this short review of possible schemes of neutrino mixing,
let us turn to the neutrino oscillations,
a phenomenon that was first considered
by B. Pontecorvo \cite{Pontecorvo}.
If there is neutrino mixing  and
neutrino masses are small,
the state of a flavor neutrino $\nu_l$ with
momentum $ |\vec{p}| \gg m_i $ is a {\em coherent superposition} of the 
states of neutrinos with  definite mass and negative helicities:
\begin{equation}
|\nu_l > = \sum_i |i><i|\nu_l>
\label{13}
\end{equation}
where $ |i> $ is an eigenstate of the free Hamiltonian,
\begin{equation}
H|i>  =  E_i|i>
\quad , \qquad
E_i \simeq {p+\frac{m_{i}^{2}}{2p}}
\qquad \mbox{and} \qquad
<\nu_l|i>=U_{li}
\;.
\label{15}
\end{equation}
If the beam of neutrinos at $t=0$ is described by the state 
$ |\nu_l> $,
at time $t$ for the state vector of the beam we have 
\begin{equation}
|\nu_l>_t =e^{-iHt}|\nu_l>
\label{16}
\end{equation}
The flavor content of the neutrino beam is analyzed with 
the help of CC weak interactions.
For the amplitude of the transition
$ \nu_l \to \nu_{l'}$
at the time $t$,
from Eqs.(\ref{13}) and (\ref{16}) we have 
\begin{equation}
A_{\nu_{l}\to \nu_{l'}}(t) = 
< \nu_{l'}|e^{-iHt}|\nu_{l} > 
= \sum_{i} U_{l'i} \, e^{-iE_{i}t} \, U^{*}_{li}  
\label{17}
\end{equation}
Analogously,
for the transition amplitude between antineutrino flavor 
states we have
\begin{equation}
A(\bar{\nu}_l \to \bar{\nu}_{l'}) =
\sum_{i} U^{*}_{l'i} \, e^{-iE_it} \, U_{li}
\label{18}
\end{equation}
Comparing the expressions (\ref{17}) and (\ref{18}),
we have the following relation
for the transition probabilities:
\begin{equation}
P({\nu_l \to \nu_{l'}}) =
P({{\bar\nu_{l'}} \to \bar\nu_l})
\label{19}
\end{equation}
where
$ P(\nu_l \to \nu_{l'})=
|{A(\nu_l\to \nu_{l'}}|^2$.
It is obvious that this relation
is a consequence of CPT invariance.

If CP invariance in the lepton sector takes place, we have
\begin{equation}
U_{li} = U^{*}_{li} \quad \mbox{for Dirac neutrinos}
\qquad \mbox{and} \qquad  
U_{li}\eta_{i} = U^{*}_{li} \quad \mbox{for Majorana neutrinos.}
\label{20}
\end{equation}
Here $\eta_i$ is
the CP parity of the Majorana neutrino with mass $m_i
(\eta_i=\pm i)$.
From Eqs.(\ref{17}), (\ref{18}) and (\ref{20})
it follows that in the case of CP invariance in the lepton 
sector 
\begin{equation}
P({\nu_l \to \nu_{l'}})=
P({\bar \nu_l\to \bar\nu_{l'}})
\label{21}
\end{equation}

Let us enumerate the neutrino masses in the following way:
\begin{equation}
m_1 < m_2 < m_3 \, ...
\label{22}
\end{equation}
The expression for the transition amplitude can be written in the form
\begin{eqnarray}
A({\nu_l \to \nu_{l'}}) =
e^{-i{E_1}t}
\left[
\sum_{i=2,3} U_{l'i}
\left(
{e^{-i\frac{\Delta {m_{i1}^2}R}{2p}}-1}
\right) U^*_{li} + \delta _{l'l}
\right]
\label{23}
\end{eqnarray}
where $\Delta m^2_{i1}=m_i^2-m^2_1$, and $ R\simeq {t} $ is the
distance between the neutrino source and the detector.
From Eq.(\ref{23})
it is obvious that transitions between different flavor neutrino states
require both mixing (non-diagonality of U)
and $\Delta m_{i1}^2\neq 0 $.
For
neutrino oscillations to be observable,
it is necessary that at least
one $ \Delta m^2 $ satisfy the following inequality:
\begin{equation}
\Delta m^2 \gtrsim \frac{p\;(\mbox{MeV})}{R\;(\mbox{m})}
\label{24}
\end{equation}
Let us notice that from this inequality it follows that solar 
neutrino experiments have an enormous sensitivity to the parameter
$ \Delta m^2 $ ($ \Delta m^2 \geq 10^{-10} \, \mbox{eV}^2 $).

The data on the search for neutrino oscillations are usually
analyzed under the simplest assumption that only two flavor neutrino 
fields are mixed. In this case
\begin{equation}
U =
\left(\begin{array}{cc}
 \cos{\theta} & \sin{\theta} \\
-\sin{\theta} & \cos{\theta}
\end{array} \right)
\label{124}
\end{equation}
where $\theta$ is the mixing angle.
From Eq.(\ref{18}),
for the transition 
probabilities we have 
\begin{eqnarray}
P({\nu_{l} \to \nu_{l'}}) & = &
\frac{1}{2}\sin^2{2\theta}
\left( 1 - \cos\frac{\Delta m^2 R}{2 p} \right)
\label{25a}
\\
P(\nu_{l}\to \nu_{l}) & = & 1 - P({\nu_{l} \to \nu_{l'}})
\label{25b}
\end{eqnarray}
The more realistic case is that with three generation mixing
and a neutrino mass hierarchy \cite{BFP92}.
Let us assume that $ \Delta m^2_{21} $
is so small that in any experiment with terrestrial or atmospheric
neutrinos
\begin{equation}
\frac{\Delta m_{21}^{2} R}{p} \ll 1.
\label{26}
\end{equation}
The mixing between the first and the second generation with small
$ \Delta m_{21}^{2} $ can be responsible for the suppression of the
flux of solar $ \nu_e$'s.
For the probability of
$ \nu_l \to \nu_{l'} $ ($l'\neq{l}$)
transitions,
from Eq.(\ref{23})
we have
\begin{equation}
P({\nu_l \to\nu_{l'}})=
\frac{1}{2} \, {A_{\nu_{l};\nu_{l'}}}
\left( 1 - \cos \frac{\Delta m^2}{2 p} \right)
\label{27}
\end{equation}
Here
\begin{equation}
A_{\nu_{l'};\nu_l} = 4 |{U_{l'3}}|^2 |{U_{l3}}|^2
\label{28}
\end{equation}
is the amplitude of
$ \nu_l \leftrightarrow \nu_{l'} $
oscillations and $\Delta m^2=m^2_3 - m^2_1 $.
The 
expression for the survival probability
can be obtained from the conservation
of the total probability. We have
\begin{equation}
P({\nu_l\to\nu_l}) =
1 - \sum_{l'} P({\nu_l\to\nu_{l'}})
=
1 - \frac{1}{2} \, B_{\nu_l;\nu_l}
\left( 1 - \cos \frac{\Delta m^2 R}{2p} \right)
\label{29}
\end{equation}
where
\begin{equation}
B_{\nu_l;\nu_l}=\sum_{l'\neq l} A_{\nu_l;\nu_l'}
\label{30}
\end{equation}
Using the unitarity of the mixing matrix,
from Eqs.(\ref{28}) and (\ref{30})
we find
\begin{equation}
B_{\nu_l;\nu_l} = 4|{U_{l3}}|^2 \left(1-|{U_{l3}|}^2\right) 
\label{31}
\end{equation}
Thus, if the inequality (\ref{26})
is satisfied for oscillations of terrestrial and/or
atmospheric neutrinos:

\noindent
1.
All the oscillation channels
$\nu_\mu \leftrightarrow {\nu_\tau}$, 
$\nu_\mu \leftrightarrow {\nu_e}$,
$\nu_e \leftrightarrow {\nu_\tau}$
are characterized by the same $ \Delta{m^2} $.

\noindent
2.
The amplitudes of exclusive and inclusive channels are connected by the 
relation (\ref{30}).

\noindent
3.
The relation (\ref{21}) is satisfied
even if CP is violated in the lepton 
sector.

\noindent
4.
The oscillations in all channels are characterized by the three parameters
$ \Delta m^2$, ${|U_{e3}|}^2$ and ${|U_{\mu 3}|}^2$
(${|U_{\tau 3}|}^2= 1-{|U_{e 3}|}^2-{|U_{\mu 3}|}^2$).

In the papers \cite{BBGK}
the existing data were analyzed in the framework of the
model with a neutrino mass hierarchy.
We shall describe the results of these
analyses later.

Now we turn to the discussion of the transitions of flavor neutrinos
in matter \cite{Wolfenstein,MSW}.
Let us consider a beam of neutrinos with momentum $\vec{p}$.
For
the wave function in the flavor representation
$ {a_{\nu_l}}(t) = <\nu_l|\psi (t)> $
the following evolution equation holds:
\begin{equation}
i\frac{\partial{a_{\nu_l}}}{\partial{t}} = H a_{\nu_l}
\qquad \qquad \mbox{with} \qquad \qquad
H = H_0 + H_I
\label{33}
\end{equation}
The free Hamiltonian $ H_0 $ is given by
\begin{equation}
<\nu_l' |H_0| \nu_l> =
\left(UEU^\dagger\right)_{l'l} =
p \, \delta _{l'l} + \left(U\,\frac{m^2}{2 p}\,U^\dagger\right)_{l'l}
\label{34}
\end{equation}
The second term of $H $ is the effective Hamiltonian of
the coherent 
interactions of the neutrino with matter.
The neutral current interaction is
$ \nu_e $-$ \nu_\mu $-$ \nu_\tau $ symmetric.
This interaction cannot 
change the flavor content of the beam.
The contribution to $ H_I $ comes 
from the CC part of the $ \nu_e $-$ e $ interaction:
\begin{equation}
H_I (t) = 2 \, \frac{G_F}{\sqrt{2}} \int{\bar\nu_{eL}}
{\gamma^\alpha} \nu_{eL} {\bar{e}} \gamma_\alpha (1 + \gamma_5) e
\, d^3 x
\label{35}
\end{equation}
Taking into account that
\begin{equation}
<\phi|\int {\bar{e}}\gamma_\alpha (1 + \gamma_5) e \, d^3 x|\phi>=
\delta_{\alpha 0} \, \rho_{e}(t)
\label{36}
\end{equation}
where $\rho_e$ is the density of electrons and $ |\phi> $ is the
state vector of matter, we have
\begin{equation}
(H_I (t))_{\nu_e;\nu_e} =
\sqrt{2} \, G_{F} \, \rho_{e}(t) \,
\delta _{\nu_e \nu_e}
\label{37}
\end{equation}
All the other matrix elements of $ H_{I} (t) $ are equal to zero.
It is important that the Hamiltonian of the effective interactions of 
neutrino with matter depends on time. It was shown
in Ref.\cite{MSW} 
that due to this
dependence resonance transitions of flavor neutrinos in matter become
possible.

Let us consider the simplest case of two neutrino flavors.
Omitting
the term $ \frac{1}{2} \mbox{Tr} H $
that is proportional to the unit matrix and does not 
change the flavor content of the beam, we obtain:
\begin{equation}
H(t) =  \frac{1}{4p}
\left( \begin{array}{cc}
-X(t) & \Delta m^2 \sin{2\theta}
\\
\Delta m^2 \sin 2\theta & X(t)
\end{array} \right)
\label{38}
\end{equation}
Here
\begin{equation}
\Delta m^2 = m^2_2 - m^2_1
\qquad \qquad \mbox{and} \qquad \qquad
X(t) = \Delta m^2\cos{2\theta} - 2\sqrt{2}G_F \rho_e (t)
\;. 
\label{39}
\end{equation}
The Hamiltonian $ H(t) $ can be easily diagonalized. We have
\begin{equation}
H(t) = U(t) E(t) U^+ (t)
\qquad \qquad \mbox{with} \qquad \qquad
U(t) =
\left( \begin{array}{cc}
\cos{\theta} (t) & \sin{\theta} (t)\\
-\sin{\theta} (t) & \cos{\theta} (t)
\end{array} \right)
\;.
\label{41}
\end{equation}
Here $ \theta (t) $
is the mixing angle in matter and
$ E_{ik} (t) = E_{i} (t) \delta _{ik} $,
$ E_{1,2} (t) $ being the energies of neutrinos in
matter (up to a constant). We have
\begin{eqnarray}
\sin{2\theta} (t) & = & \frac{\Delta m^2 \sin{2\theta}}
{\sqrt{X^2 (t) + \Delta m^4 \sin^2{2\theta}}}
\label{42a}
\\
\cos{2\theta(t)} & = & \frac{X(t)}
{\sqrt{X^2 (t) + \Delta m^4 \sin^2{2\theta}}}
\label{42b}
\\
E_{1,2} & = & \mp \frac{1}{4p} \sqrt{X^2 (t) + \Delta m^4 \sin^2{2\theta}}
\label{42c}
\end{eqnarray}
As it is clear from Eqs.(\ref{42a}) and (\ref{42b}),
the mixing angle in matter depends on the 
density of electrons.
Assume that at some point $x_R = t_R $ the
condition 
\begin{equation}
\Delta m^2 \cos{2\theta} = 2\sqrt{2} \, G_F \, \rho_e(t_R) \, p 
\label{43}
\end{equation}
is
satisfied.
At this point the diagonal elements of the Hamiltonian $ H(t) $ 
are equal to zero and,
as it follows from Eq.(\ref{42c}),
for any value of
$ \theta \neq 0 $ the mixing is maximal:
$ \theta (t_R) = \pi/4 $.

The condition (\ref{43}) is called resonance condition.
Let us 
notice that at the point $ t=t_R $
the distance between the energy levels of
neutrinos in matter is minimal:
\begin{equation}
E_2 (t_R) - E_1 (t_R) = \frac{\Delta m^2 \sin{2\theta}}{2p}
\label{44}
\end{equation}
The resonance condition (\ref{43}) can be written in the form 
\begin{equation}
\Delta m^2\cos{2\theta} \simeq
0.7 \times 10^{-7}\rho E \, \mbox{eV}^2
\label{45}
\end{equation}
where $ \rho $ is the density of matter in g/cm$^3$.
In the center of the sun
$ \rho \simeq 10^2 \, \mbox{g}/\mbox{cm}^3 $
and
the energy
of solar neutrinos is $ \simeq 1 $ MeV.
Thus, for solar neutrinos the resonance condition (\ref{45})
is satisfied at
$ \Delta m^2 \simeq 10^5 \, \mbox{eV}^2 $.
The solution of the evolution equation 
shows that in a wide region of the values of the parameters
$\Delta m^2 $
and $ \sin^2{2\theta} $
the probability of solar neutrinos to survive 
depends on neutrino energy and can be significantly less than one.

Now we turn to a short discussion of the results of the experiments
aimed
to reveal the effects of neutrino masses and mixing
\cite{Winter}.
There are three groups
of experiments of this type:

\noindent
{\bf 1}.
The experiments on the search for effects of neutrino mass through 
precise measurement of the high energy part of the $\beta$-spectrum.
The classical
process is $\beta$-decay of $^3H$:
\begin{equation}
{^3} H \to {^3}He + e^{-} + \bar {\nu}_e
\label{46}
\end{equation}
The spectrum of the electrons in this decay is determined by the phase
space
\begin{equation}
\frac{dN}{dT} = CpE(Q-T)\sqrt{(Q-T)^2 - m_{\nu}^2} \, F(E)
\label{47}
\end{equation}
Here $p $ and $E $ are electron momentum and total energy,
$T=E-m_e$,
$m_\nu$ is the mass of the electron neutrino,
$F(E)$ is the Fermi-function
that describes the electromagnetic interaction of the final particles,
and
$ Q \simeq 18.6 \, \mbox{keV} $ is the released energy.
There is no indications in
favor of non-zero $ m_\nu $ from experiments of this type.
In the 
latest experiments the following upper 
bounds were obtained:     
\begin{center}
\begin{tabular}{lll}
Mainz \hskip0.5cm \null &
$ m_\nu < 7.2 \, \mbox{eV} $ \hskip0.5cm \null &
$ (m^2_\nu = - 39 \pm 34 \pm 15 \, \mbox{eV}^2) $
\\
Troizk \hskip0.5cm \null &
$ m_\nu < 4.35 \, \mbox{eV} $ \hskip0.5cm \null &
$ (m^2_\nu = - 4.1 \pm 10.3 \, \mbox{eV}^2) $
\end{tabular}
\end{center}
Let us notice that 
$ m_{\nu_{\mu}}<160$ keV  and $ m_{\nu_{\tau}}<24 $ MeV.

\noindent
{\bf 2}.
The experiments on the search for neutrinoless double $\beta$-decay 
\begin{equation}
(A,Z) \to (A,Z+2) + e^{-} + e^{-}
\label{48}
\end{equation}
These decays are allowed only if neutrinos are massive and Majorana 
particles (the total lepton number L is not conserved).
The matrix element of 
the $ (\beta\beta)_{0\nu} $ decay is proportional to
\begin{equation}
\langle m \rangle = \sum_{i} U^2 _{ei} \, m_i
\;,
\label{49}
\end{equation}
where the factor $ U^2_{ei} $ is due to two $e$-$\nu_i$ vertices
and $m_i$ is due to the neutrino propagator.
Neutrinoless double $\beta$-decay
was not observed in experiment.
The best lower limit on the life time was
reached in the Heidelberg-Moscow experiment for ${^{76}} Ge $:
$T_{1/2} ({^{76}} Ge)\geq 7.2\times 10^{24}$ y.
From this data it 
follows that $ |\langle m \rangle| \lesssim 1 \, \mbox{eV} $.

\begin{table}[t]
\begin {center}
Table 1 \\ \medskip
\begin{tabular}{|c|c|c|} 
\hline
Reaction & Neutrino energy & Expected flux \\
& (MeV) & ($\mbox{cm}^{-2}\mbox{sec}^{-1}$)
\\ \hline
$ p+p \to d + e^{+} + \nu_e $ &
$ \leq 0.42 $ &
$ 6.0 \times {10^{10}} $ \\
$^{7}Be + e^{-} \to {^{7}Li + \nu_e}$ 
& 0.86 & $ 4.9 \times {10^{9}} $ \\
$^{8}B \to {^{8}Be + e^{+}\nu_e} $ &
$ \leq 14 $ & $5.7 \times {10^{6}}$ \\
\hline
\end{tabular}
\end{center}
\end{table}
 
\noindent
{\bf 3}.
The experiments on the search for neutrino oscillations.
The search for neutrino oscillations is the most sensitive method to reveal
neutrino masses and mixing.
We shall considered first the solar neutrino
experiments.

The most important reactions of the solar $pp$ cycle in which 
neutrinos are produced presented in Table 1.
The expected fluxes are result of calculations in the framework of the 
standard solar model \cite{Bahcall} (SSM).
There is one model independent constraint 
on the fluxes  of solar neutrinos.
The energy of the sun is produced in the
transition
\begin{equation}
2 e^- + 4 p \to {^{4}He} + 2\nu_e
\label{50}
\end{equation}
Thus the production of energy in the sun is accompanied by
the emission of neutrinos.
If we assume that the sun is in a stable state,
we have
\begin{eqnarray}
\frac{1}{2} \, Q \sum_{i=pp,...}
\left(1 - 2 \frac{\bar{E_i}}{Q}\right) \Phi_i =
\frac{L_\odot}{4 \pi R^2}
\label{51}
\end{eqnarray}
Here
\begin{eqnarray}
Q = 4 m_p + 2 m_e - m_{^{4} He} \simeq 26.7 \mbox{MeV}
\;,
\label{52}
\end{eqnarray}
$L_{\odot}$ is the luminosity of the sun,
R is the distance between the
sun and the earth, $\Phi_i $
is the total flux of neutrinos from the source $i$
($i=pp, {^{7} Be},...$)
and $\bar{E_i}$ is the average energy of neutrinos from the source i.
The 
results of the four solar neutrino experiments are presented in Table 2.
These results are presented in SNU
($ 1 \, \mbox{SNU} =
10^{- 36} \, \mbox{events} / ( \mbox{atom} \cdot \mbox{sec} ) $).
As it seen from Table 2, 
the event rates in all
solar neutrino experiments are significantly less than
the predicted event rates.
If we accept the neutrino fluxes predicted by the SSM,
the existing data can be 
described under the  simplest assumption of transitions between two 
neutrino types.
If matter effects are important,
the following values are found
for the mixing parameters:
\begin{eqnarray}
1. \hskip1cm \null &
\sin^2{2\theta} \simeq {8\times{10^{-3}}}
\quad , \qquad &
\Delta m^2 \simeq {5\times{10^{-6}}} \, \mbox{eV}^2
\quad ,
\label{53a}
\\
2. \hskip1cm \null &
\sin^2{2\theta} \simeq {0.8}
\quad , \qquad &
\Delta m^2 \simeq {10^{-5}}  \, \mbox{eV}^2
\quad .
\label{53b}
\end{eqnarray}
The existing data can also be described by vacuum oscillations.
In this 
case, for the parameters $\sin^2{2\theta}$ and $\Delta m^2$ the
following values were found: $\sin^2{2\theta} \simeq {0.8}, \Delta m^2
\simeq {8\times{10^{-11}}} \, \mbox{eV}^2 $.
 
\begin{table}[t]
\begin{center}
Table 2. The result of solar neutrino experiments.
\\ \medskip
\begin{tabular}{|c|c|c|}
\hline
Experiment, & Data & Prediction of  \\
reaction, threshold &  & SSM \\
\hline
Homestake & $2.55 \pm 0.35$ SNU & $9.3 \pm 1.4$ SNU\\
$\nu_e{^{37}{Cl}} \to {e^-{^{37}{Ar}}}, \, E_{th}=0.81 \mbox{MeV}$ & & \\
Gallex & $ 77 \pm 8.5 \pm 5 $ SNU & $ 131.5 \pm 6$ SNU\\
$\nu_e{^{71}{Ga}} \to {e^-{^{71}{Ge}}}, \, E_{th}=0.23 \mbox{MeV} $ & & \\
Sage &  $69 \pm 11 \pm 6$ SNU & $ 131.5 \pm 6$ SNU\\
Kamiokande & data/SSM = &  \\
$ \nu e \to \nu e , \, E_{th} \simeq {7 \mbox{MeV}} $ &
$ 0.51 \pm 0.04 \pm 0.06 $ & \\
\hline
\end{tabular}
\end{center}
\end{table}

The lower bound of the event rate in gallium experiments
$Q_{Ga}$ can be found from the luminosity constraint (\ref{51}).
In fact,
assuming that $P(\nu_e \to \nu_e)=1 $, we have
\begin{equation}
Q_{Ga} =
\sum_{i} \bar\sigma_i \Phi_i
\ge \bar\sigma_{pp}
\sum_{i} \Phi_i \simeq 80 \, \mbox{SNU}
\label{54}
\end{equation}
This lower bound does not contradict the values of $ Q_{Ga} $ 
measured in the Gallex and Sage experiments.
Let us compare, however, the 
data of different experiments under the  assumption that nothing
happens with
solar $ \nu_e $'s on their way from the sun to the earth.
Let us compare, for
example, the data from the Homestake and Kamiokande experiments.
In the Kamiokande experiment only
$^8{B}$ neutrinos are detected
(the threshold is $\simeq 7$ MeV).
The flux $\Phi_B$ can be determined from
the data of this experiment .
Using
this flux it is possible to obtain the contribution of $^7Be$ and other
neutrinos to the event rate of the Homestake experiment.
In Ref.\cite{Bahcall94} it was
found that
\begin{equation}
Q_{Cl}(\nu_{Be},...) = - 0.66 \pm 0.52 \, \mbox{SNU}
\label{55}
\end{equation}
From this value it follows that
\begin{equation}
Q_{Cl} (\nu_{Be},...) < 0.46 \, \mbox{SNU} \quad (95\% \mbox{C.L.})
\label{56}
\end{equation}
On the other hand, all standard solar models give
\begin{equation}
Q_{Cl}(\nu_{Be} ,...) = 1.1 \pm 0.1 \, \mbox{SNU}
\label{57}
\end{equation}
Thus,
rather strong indications in favor of neutrino
mixing follow
from the analysis of the data of different solar
neutrino experiments.

If the mass of the heaviest neutrino is in the eV region,
neutrinos can
solve the problem (or part of the problem) of dark matter.
Two new
experiments at CERN,
CHORUS \cite{CHORUS} and NOMAD \cite{NOMAD},
are searching
for $\nu_\mu \leftrightarrow {\nu_\tau}$ oscillations.
For $\Delta m^2 \ge 10 \, \mbox{eV}^2$
these experiments are sensitive to
$ A_{\nu_{\mu};\nu_{\tau}} \ge 6 \times 10^{-4} $. 
If there is a hierarchy of neutrino masses and
$ \Delta m^2_{21} = \Delta m^2 _2 -m^2 _1 $
is relevant for the suppression of the flux
of solar $\nu_{e}$'s,
the probability of
$ \nu_l \to \nu_{l'} $ ($l'\neq l$)
transitions
is given by Eq.(\ref{27}).
As it is well known,
there is a hierarchy of couplings between generations in the quark
sector.
Let us assume that the hierarchy of couplings is a general phenomenon
valid also for the lepton sector.
In this case
\begin{equation}
|U_{e3}|^2 \ll |U_{\mu3}|^2 \ll 1
\qquad \qquad \mbox{and} \qquad \qquad
|U_{\tau 3}|^2 \simeq 1
\;.
\label{58}
\end{equation}
For the oscillation amplitudes,
from Eqs.(\ref{28}) and(\ref{58}) we obtain
\begin{eqnarray}
A_{\nu_{e};\nu_{\tau}} \ll {A_{\nu_{\mu};\nu_{\tau}}} \ll 1
\label{59a}
\\
A_{\nu_{\mu};\nu_{e}} \simeq
{\frac{1}{4} A_{\nu_{e};\nu_{\tau}} A_{\nu_{\mu};\nu_{\tau}}}
\label{59b}
\end{eqnarray}
Thus, if there is a hierarchy of couplings in the lepton sector,
$ \nu_{\mu} \to \nu_{\tau} $ is the dominant transition.
 
In conclusion,
we shall discuss some indications in favor of
neutrino mixing that follow from the data on beam-stop neutrino 
experiment and atmospheric neutrino experiments.
Recently the LSND collaboration published
\cite{LSND}
the results of the experiment
on search for $ \bar{\nu}_{\mu} \to \bar{\nu}_{e} $ oscillations.
The sources of neutrinos in this experiment are
$ \pi^{+} \to \mu^{+} \nu_{\mu} $
and
$ \mu^{+} \to e^{+} \nu_{e} \bar\nu_{\mu} $
decays at rest.
In the experiment $\bar{\nu}_{e}$'s were searched 
for via the observation of the process
$ \bar{\nu}_e p \to e^{+} n $.
Nine
candidate events were found with
an estimated background of $2.1\pm 0.3$ events.
A possible interpretation of this result (which requires confirmation)
are
$\bar{\nu}_{\mu} \to {\bar{\nu}_e}$
oscillations with an
amplitude $ 10^{-3} \le {A_{\nu_{\mu};\nu_e}} \le 10^{-2}$
and
$ 2 \times 10^{-1} \, \mbox{eV}^2 \le
{\Delta m^2 } \le
{5 eV^2} \, \mbox{eV}^2 $. 
We have analyzed the LSND result
together with the results of the other experiments
in the framework of the model with mixing of three massive neutrino
fields and a neutrino mass hierarchy \cite{BBGK}.
If the elements $|U_{e3}|$ 
and  $|U_{\mu3}|$ are small
(as in the case of a hierarchy of couplings in
the lepton sector),
the LSND positive signal contradicts the negative
results from the other experiments.
The LSND  result is compatible with
the results from other experiments only in the case of
a rather unusual mixing
in the lepton sector with large $|U_{\mu3}|$ and small $|U_{e3}|$ and 
$|U_{\tau3}|$
(in this case $\nu_\mu$ is the heaviest neutrino).

Atmospheric  neutrinos are produced in the decays 
\begin{equation}
\pi \, (K) \to \mu \, \nu_{\mu}
\quad , \qquad \qquad
\mu \to e \, \nu_{e} \, \nu_{\mu}
\label{60}
\end{equation}
Thus,
in the atmospheric neutrino flux $N_{\nu_{\mu}}/N_{\nu_e}\simeq2$.
For the ratio
\begin{equation}
R =
\left( \frac{N_\mu}{N_e} \right)_{\mathrm{obs}}
\Bigg/
\left( \frac{N_{\mu}}{N_e} \right)_{\mathrm{MC}} 
\label{61}
\end{equation}
the following value was obtained
in the Kamiokande experiment:
\begin{equation}
R = 0.60 \, ^{+0.06}_{-0.05} \pm 0.05
\label{62}
\end{equation}
Here $ N_{\mu} (N_e ) $ is the number of muon and electron events
and
$ (N_{\mu}/N_e)_{\mathrm{MC}} $ is the predicted ratio.
The
atmospheric neutrino anomaly was observed also in
the IMB and Soudan experiments:
\begin{eqnarray}
R = 0.54 \pm 0.05 \pm 0.12
& \null \hskip1cm \null & \mbox{(IMB)}
\label{63a}
\\
R = 0.64 \pm 0.17 \pm 0.09
& \null \hskip1cm \null & \mbox{(Soudan)}
\label{63b}
\end{eqnarray}
On the other hand,
in the Frejus experiment $ R = 0.99 \pm 0.13 \pm 0.08 $.
The Kamiokande data
can be described under the assumption of
$\nu_{\mu} \leftrightarrow \nu_{\tau} $
or
$ \nu_{\mu} \leftrightarrow \nu_e $
oscillations.
For
the oscillation parameters the following values were obtained:
\arraycolsep=0pt
\begin{eqnarray}
5 \times 10^{-3}
\le
\Delta m^2
\le
3 \times 10^{-2} \, \mbox{eV}^2
\quad , \quad & \quad
0.7
\le
\sin^2 2\vartheta
\le
1
\qquad & \quad
( \nu_{\mu} \leftrightarrow \nu_{\tau} )
\label{64}
\\
7 \times 10^{-3}
\le
\Delta m^2
\le
8 \times 10^{-2} \, \mbox{eV}^2
\quad , \quad & \quad
0.6
\le
\sin^2 2\vartheta
\le
1
\qquad & \quad
( \nu_{\mu} \leftrightarrow \nu_{e} )
\label{65}
\end{eqnarray}

New experiments in search for neutrino oscillations are now under
preparation.
I have in mind the so called long baseline neutrino experiments:
\begin{center}
\begin{tabular}{ll}
KEK--Super-Kamiokande & (250 Km)
\\
Fermilab--Soudan & (730 Km)
\\
CERN--Gran Sasso & (730 Km)
\end{tabular}
\end{center}
The appearance
($\nu_{\mu} \to \nu_{\tau}$,
$\nu_{\mu} \to \nu_e$)
and disappearance
($\nu_{\mu} \to {\nu_{\mu}}$)
channels will be investigated
and the indications in favor of neutrino mixing
that come from the atmospheric neutrino experiments will be checked.

In conclusion, we would like to stress that the problem of neutrino
masses and mixing is the central problem of today's neutrino physics.
The 
investigations of this problem could allow us to reach physics beyond the 
standard model.
At present different indications exist that neutrinos are 
massive and mixed.
New experiments may appear crucial for the problem.

\end{document}